\documentclass[twocolumn,letter]{jpsj2}
\usepackage{bm}
\usepackage{color}
\usepackage{txfonts}
\title{Parity-Mixed Superconductivity in Locally Non-centrosymmetric System}

\author{Tomohiro YOSHIDA$^1$, Manfred SIGRIST$^2$, and Youichi YANASE$^1$}
\inst{$^1$Department of Physics, Niigata University, Niigata 950-2181, Japan \\
$^2$Theoretische Physik, ETH-Zurich, 8093 Zurich, Switzerland}

\abst{We study the parity-mixed superconductivity in locally non-centrosymmetric systems. 
In multilayer systems an inhomogeneous Rashba spin-orbit coupling is induced by the local violation of inversion symmetry. 
Our previous study revealed that a pair-density wave (PDW) phase, with a sign-modulated order parameter, is stabilized in the spin-singlet multilayer superconductors owing to the spin-orbit coupling. 
In this letter, we show that the uniform spin-triplet superconductivity emerges through parity mixing in the PDW phase, taking into account a weak interaction in the spin-triplet channel, which was neglected in our previous study.
The spin-triplet superconducting phase is nonunitary owing to field-induced parity mixing. 
The critical magnetic field is markedly increased by the emergence of spin-triplet superconductivity.  
We calculate the density of states and analyze the signature specific to this phase. 
}

\kword{locally non-centrosymmetric superconductors, staggered Rashba spin-orbit coupling, parity mixing, spin-triplet superconductivity}

\begin{document}
\maketitle
Spin-orbit coupling arises from relativistic effects and is responsible for many intriguing features in condensed matter physics.
For example, the spin-Hall effect~\cite{Science.301.1348,PRL.92.126603}, chiral magnetism~\cite{Nature.465.901,PRL.108.107202}, and topologically insulating and superconducting phases,~\cite{RevModPhys.82.3045,RevModPhys.83.1057} as well as non-centrosymmetric superconductivity~\cite{NCSC} attract continued interest. 
Recent studies of ``locally'' non-centrosymmetric systems (LNCSs) have led to the introduction of a novel class of superconducting states affected by spin-orbit coupling in a particular manner.~\cite{JPSJ.79.084701,PRB.84.184533,JPSJ.81.034702,PRB.85.220505,PRB.86.100507,PRB.86.134514,JPSJ.82.074714} 
Simple crystal structures displaying local non-centrosymmetricity are realized in various multilayer systems, such as the recently fabricated artificial superlattices of CeCoIn$_{5}$~\cite{NatPhys.7.849} and multilayer high-$T_{\rm c}$ cuprates.~\cite{PRL.96.087001} 
In multilayer systems, the mirror symmetry is broken for each layer except in center layers; therefore, a layer-dependent Rashba spin-orbit coupling emerges.~\cite{JPSJ.81.034702} 
When the strength of spin-orbit coupling is comparable to (or larger than) that of interlayer coupling, the effects of broken local inversion symmetry manifest in various superconducting properties,~\cite{JPSJ.81.034702} and even rather exotic superconducting phases appear in the magnetic field.~\cite{PRB.86.134514,JPSJ.82.074714} 
For instance, the pair-density wave (PDW) phase, where the order parameter switches sign between layers, is stabilized in a magnetic field perpendicular to the conducting planes,~\cite{PRB.86.134514} while the complex-stripe phase is stable for an in-plane magnetic field.~\cite{JPSJ.82.074714}

In the context of PDW states, it is of interest to analyze the aspect of parity mixing more deeply, i.e., the feature that spin-orbit coupling admixes an odd-parity spin-triplet ``$p$-wave'' order parameter to a dominant even-parity spin-singlet $s$-wave superconducting phase.~\cite{PRB.86.134514,JPSJ.82.074714}
It is straightforward to see that, in a bilayer system, the even-parity $s$-wave PDW phase (alternating sign from layer to layer) is accompanied by a uniform odd-parity component. 
Therefore, formally, we might consider this PDW phase among the odd-parity spin-triplet superconducting phase, although the dominant condensation energy gain occurs through the even-parity part. 
Thus, we can say that spin-triplet superconductivity is stabilized by the joint forces of spin-orbit coupling and a magnetic field. 
The purpose of this study is to analyze this superconducting state in locally non-centrosymmetric multilayer systems.

Interestingly, we find that the order parameter of induced spin-triplet superconductivity is nonunitary. 
Although nonunitary phases are unstable in purely spin-triplet superconductors~\cite{RevModPhys.47.331,RevModPhys.63.239}, they are induced by field-induced parity mixing, which originates from the synergy of spin-orbit coupling and a magnetic field in the LNCS.
It is shown that the induced nonunitary spin-triplet pairing significantly enhances the critical magnetic field in a characteristic way.
Moreover, we propose an experimental test by searching for a specific signature of parity-mixed superconducting phases in the local density of states.

In our study, we adopt a quasi-two-dimensional multilayer model taking into account layer-dependent Rashba spin-orbit coupling and the parity mixing of $s$-wave and $p$-wave superconductivities:
\begin{eqnarray}
  {\cal H}&=&\sum_{{\bm k},s,m}\xi({\bm k})c^\dagger_{{\bm k}sm}c_{{\bm k}sm}+t_\perp \sum_{{\bm k},s,\langle m,m'\rangle} c^\dagger_{{\bm k}sm}c_{{\bm k}sm'} \nonumber \\
  &&+\sum_{{\bm k},s,s',m}\alpha_m {\bm g}({\bm k})\cdot {\bm \sigma}_{ss'}c^\dagger_{{\bm k}sm}c_{{\bm k}s'm}  \nonumber \\
  &&-\mu_{\rm B}H\sum_{{\bm k},s,m}sc^\dagger_{{\bm k}sm}c_{{\bm k}sm} \nonumber \\
  &&+\sum_{{\bm k},{\bm k}',s,s',m}V_{ss'}({\bm k},{\bm k}')c^\dagger_{{\bm k}s m}c^\dagger_{-{\bm k}s' m}c_{-{\bm k}'s' m}c_{{\bm k}'s m}, \nonumber \\
  \label{eq1}
\end{eqnarray}
where $c^\dagger_{{\bm k}sm}$ ($c_{{\bm k}sm}$) is the creation (annihilation) operator for an electron with momentum ${\bm k}$ and spin $s$ on layer $m$.
We assume the nearest-neighbor hopping tight binding form $\xi({\bm k})=-2t(\cos k_x+\cos k_y)-\mu$ on a square lattice, and a small interlayer hopping $t_\perp/t=0.1$. 
We choose the chemical potential $\mu/t=2$ leading to the electron number density per site $n = 1.60$, unless explicitly mentioned otherwise. 
Later, we will discuss the electron density dependence of the superconducting phase. 
It is shown that the results are almost independent of the density except around half-filling $n=1$, namely, $\mu=0$. 
By symmetry, we can choose the $g$-vector for the spin-orbit coupling of the Rashba type,~\cite{SovPhysSolidState.1.368} ${\bm g}({\bm k})=(-\sin k_y,\sin k_x,0)$.
The coupling constants $\alpha_m$ depend on the layer and are antisymmetric with respect to the reflection at the center of the multilayer structure, so that the global inversion symmetry is conserved. 
For instance, $(\alpha_1,\alpha_2)=(\alpha,-\alpha)$ for bilayer systems and $(\alpha_1,\alpha_2,\alpha_3)=(\alpha,0,-\alpha)$ for trilayer systems.
We assume $\alpha/t_\perp=3$ throughout this paper. 
Since we consider superconductors in the paramagnetic limiting regime with a large Maki parameter, such as an artificially grown superlattice of CeCoIn$_5$~\cite{NatPhys.7.849} and some multilayered high-$T_{\rm c}$ cuprates,~\cite{PRL.96.087001} we take into account the paramagnetic depairing effect through the Zeeman coupling term and neglect the orbital depairing effect. 
Note, that it is sufficient, for our purpose, to consider one superlattice unit cell along the out-of-plane direction.

As mentioned above, a key issue of this study is the analysis of parity mixing in Cooper pairing.
To this goal, we assume attractive interaction in both the $s$-wave and $p$-wave channels, $V_{ss'}({\bm k},{\bm k}')=-V_{\rm s}\delta_{s,-s'}-2V_{\rm t}(\sin k_x \sin k_x'+\sin k_y\sin k_y')$, where $V_{\rm s}$ and $V_{\rm t}$ denote the coupling constant of $s$-wave and $p$-wave attractive interactions, respectively.
In the following, we fix $V_{\rm s}/t=1.7$ and vary the parameter $V_{\rm t}/t$. 
The unit of energy is chosen as $t=1$. 
We analyze the model on the basis of the Bogoliubov$-$de Gennes (BdG) equation, and calculate the layer-dependent order parameters with mixed parity: $\Delta_{ss'}^m({\bm k})=-\sum_{{\bm k}'}V_{ss'}({\bm k},{\bm k}')\langle c_{-{\bm k}'s'm}c_{{\bm k}'sm}\rangle$. 
Using the conventional notation, order parameters are represented as $\hat{\Delta}_m({\bm k})=[\psi_m \hat{\sigma}_0+{\bm d}_m({\bm k})\cdot \hat{{\bm \sigma}}]{\rm i} \hat{\sigma}_y$, where  $\psi_m$ is the scalar order parameter in the spin-singlet channel and the so-called $d$-vector ${\bm d}_m$ denotes the vector order parameter in the spin-triplet channel. 
In order to obtain phase diagrams against temperature and magnetic field, we calculate the free energy of several metastable states and determine the stable superconducting phase.

\begin{figure}[htbp]
  \begin{center}
    \includegraphics[width=70mm]{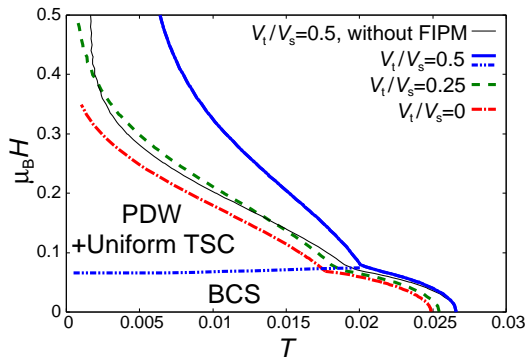}
    \caption{(Color online) Upper critical fields for several values of $p$-wave attractive interaction $V_{\rm t}$ (bilayers). 
      We fix the $s$-wave attractive interaction $V_{\rm s}=1.7$ and assume $V_{\rm t}/V_{\rm s} = 0$ (dash-dotted line), $0.25$ (dashed line), and $0.5$ (solid line). 
      The first-order phase transition line separating the BCS and PDW phases is depicted for $V_{\rm t}/V_{\rm s} = 0.5$ (dash-two-dotted line). 
      In the PDW phase, a uniform spin-triplet superconductivity is induced. [See Fig.~\ref{fig2}(a).] 
      For comparison, the upper critical field without taking the field-induced parity mixing into account is shown by the thin solid line. 
    }
    \label{fig1}
  \end{center}
\end{figure}
First, we study the bilayer system. 
Figure~\ref{fig1} shows its $T$-$H$ phase diagrams.
We see that the critical temperature $T_{\rm c}$ is slightly enhanced by the $p$-wave attractive interaction $V_{\rm t}$ at zero magnetic field. On the other hand, the upper critical field $H_{\rm c2}$ markedly increases with $V_{\rm t}$, although the $p$-wave attractive interaction $V_{\rm t}$ is small. 
Indeed, the largest $V_{\rm t} =0.85$ ($V_{\rm t}/V_{\rm s} = 0.5$) in Fig.~\ref{fig1} yields a very low transition temperature $T_{\rm c} < 6 \times 10^{-4}$ in the absence of the $s$-wave attractive interaction, namely, at $V_{\rm s}=0$. 
This means that superconductivity is mainly caused by spin-singlet pairing, but that the upper critical field is significantly enhanced by parity mixing through the spin-triplet component. 
Thus, even weak parity mixing renders locally non-centrosymmetric superconductors more robust against paramagnetic depairing effects.

In Fig.~\ref{fig1}, we show the first-order transition line in the superconducting state for $V_{\rm t}/V_{\rm s}=0.5$. 
Since global inversion symmetry is conserved in the presence of the staggered Rashba spin-orbit coupling, the superconducting states are classified on the basis of the parity.
In the low-magnetic-field region (BCS phase), the order parameter of spin-singlet pairing is uniform, $(\psi_1,\psi_2) = (\psi,\psi)$, and a staggered order parameter in the spin-triplet channel, namely, $( {\bm d}_1, {\bm d}_2 ) = ( {\bm d}, -{\bm d} )$, is induced by spin-orbit coupling. 
The parity of both components is even.
On the other hand, the spin-singlet order parameter is staggered, $(\psi_1,\psi_2) = (\psi,-\psi)$, and therefore the parity is odd in the high-magnetic-field region.~\cite{PRB.86.134514}
Adopting the definition given in Ref.~\citen{PRL.102.207004}, we call this state the pair-density wave (PDW) state, since the order parameter modulates the length scales of the crystal lattice constant. 
Originally, the PDW state in the spin-triplet channel was studied in a phenomenological manner.~\cite{PRL.102.207004} 
On the other hand, the PDW state in the spin-singlet channel is stabilized in our model.
Interestingly, a uniform order parameter of spin-triplet pairing with an odd parity, $( {\bm d}_1, {\bm d}_2 ) = ( {\bm d}, {\bm d} )$, is induced by staggered Rashba spin-orbit coupling in the odd parity PDW phase. 
We would like to stress that we are not required to assume a substantial attractive interaction in the spin-triplet channel in order to stabilize the PDW phase. 
Indeed, the condensation energy mainly comes from spin-singlet pairing, and the cost of Josephson coupling energy due to the sign change of the spin-singlet order parameter is compensated for by the magnetic energy gained in the PDW state~\cite{PRB.86.134514}. 
Then, spin-orbit coupling induces a uniform spin-triplet pairing.

\begin{figure}[htbp]
  \begin{center}
    \begin{tabular}{cc}
      \begin{minipage}{0.5\hsize}
        \includegraphics[width=40mm]{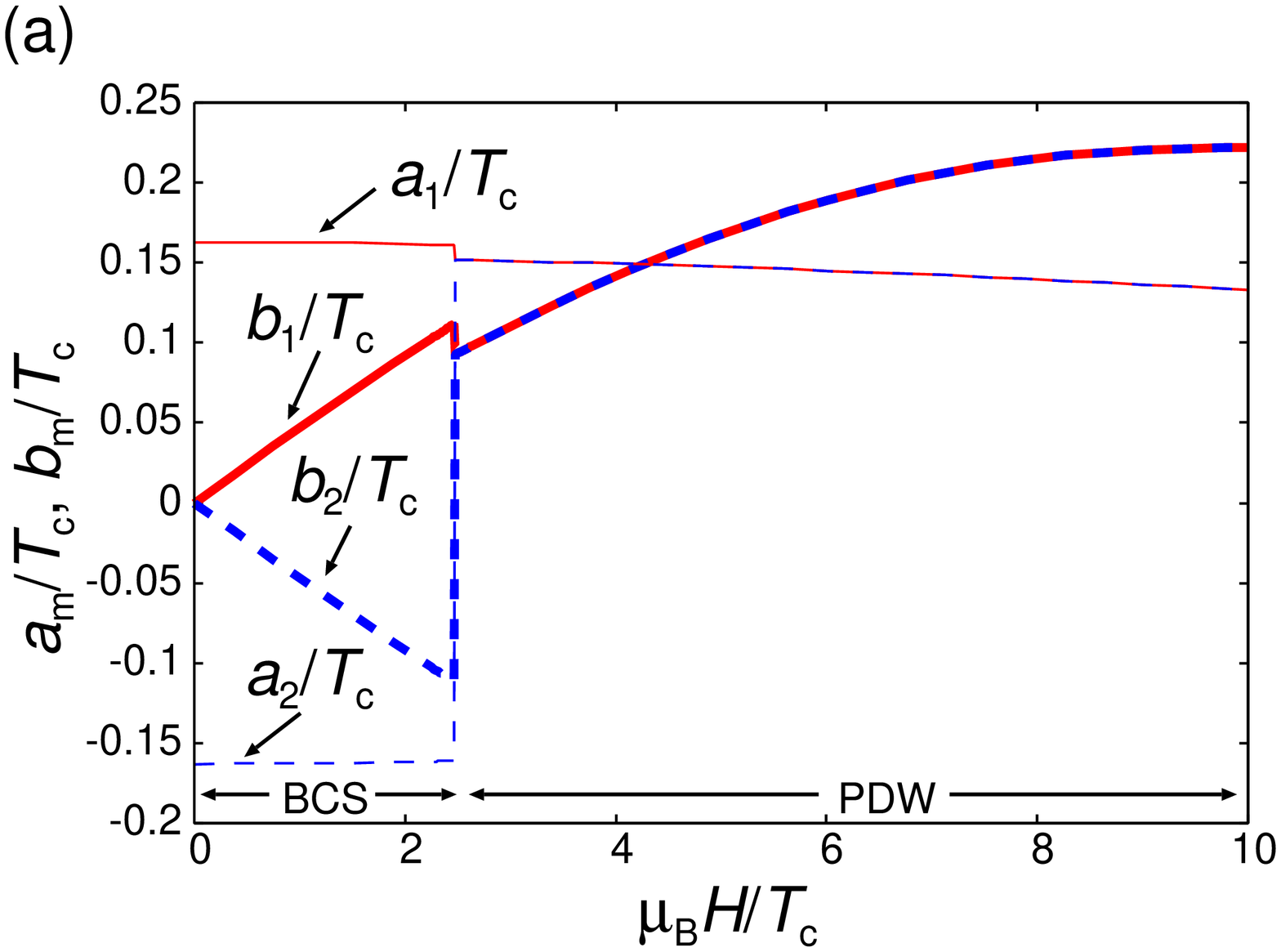}
      \end{minipage}
      \begin{minipage}{0.5\hsize}
        \includegraphics[width=40mm]{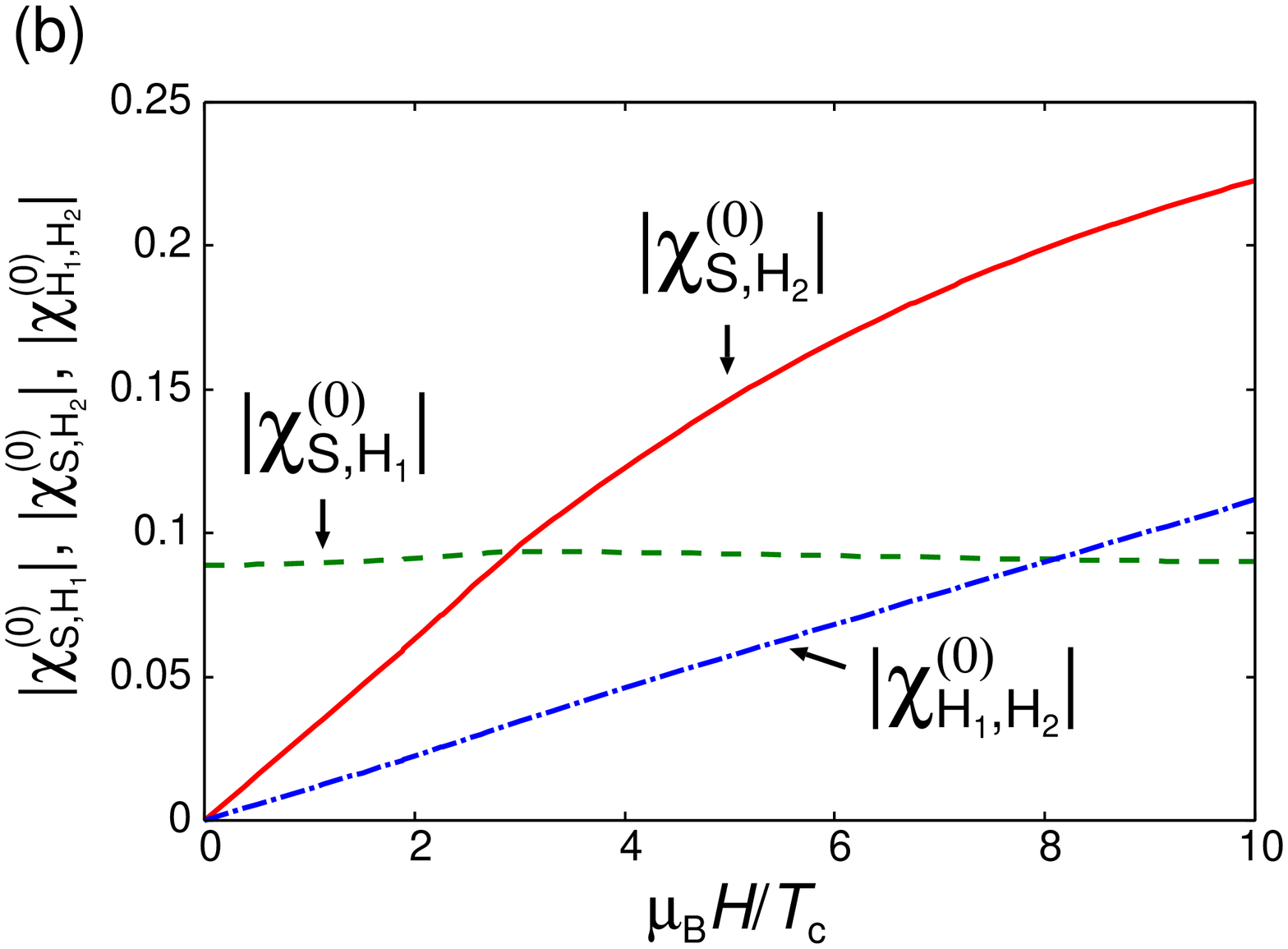}
      \end{minipage}
    \end{tabular}
    \caption{(Color online) (a) Magnetic field dependence of the $d$-vector for $V_{\rm t}/V_{\rm s}=0.5$ (bilayers). 
      We show the coefficients $a_m$ and $b_m$ in Eq.~(\ref{eq2}) at a low temperature $T/T_{\rm c}$=0.0959.  
      The thin and thick lines show $a_m$ and $b_m$, respectively. 
      The solid and dashed lines are shown for $m=1$ and $2$, respectively. 
      (b) Amplitude of irreducible superconducting susceptibility. 
      We show the magnetic field dependence at the transition temperature $T = T_{\rm c}(H)$ for $V_{\rm t}/V_{\rm s}=0.5$.
      The solid and dash-dotted lines show the field-induced components $|\, \chi^{(0)}_{\rm S,H_2}|$ and $|\,\chi^{(0)}_{\rm H_1,H_2}|$, respectively. 
      The dashed line shows $|\, \chi^{(0)}_{\rm S,H_1}|$ yielding parity mixing at zero magnetic field.
    } 
    \label{fig2}
  \end{center}
\end{figure}
We now characterize the $d$-vector of spin-triplet superconductivity induced by parity mixing. 
Our BdG calculation reveals, for the $d$-vector, a structure of the form 
\begin{eqnarray}
  {\bm d}_m({\bm k})&=&a_m(-\sin k_y\hat{x}+\sin k_x\hat{y})   \nonumber \\
  &&+{\rm i}b_m(\sin k_x \hat{x}+\sin k_y\hat{y}).
  \label{eq2}
\end{eqnarray}
As in the Rashba-type non-centrosymmetric superconductor~\cite{NCSC,PRL.92.097001,JPSJ.76.051008,JPSJ.76.124709}, the $p$-wave order parameter ${\bm d}_m({\bm k})=a_m(-\sin k_y\hat{x}+\sin k_x\hat{y})$ (hereafter, called the helical-1 state) is induced at zero magnetic field. 
On the other hand, another component represented by the coefficient $b_m$ (called the helical-2 state) is induced at finite magnetic fields; therefore, the spin-triplet pairing is nonunitary, i.e., ${\bm d}_m({\bm k})\times {\bm d}_m^\ast({\bm k}) \ne 0$, in the time reversal channel. 
Note that mixed-parity states are also nonunitary in the inversion channel. 
As shown in Fig.~\ref{fig2}(a), both components are staggered in the BCS phase as $a_1 = - a_2$ and $b_1 = - b_2$, while they are uniform in the PDW phase, $a_1 = a_2$ and $b_1 = b_2$, as mentioned above.

In Fig.~\ref{fig2}(a), we see that the nonunitarity represented by $K_m \equiv 2 a_m b_m/(a_m^2 + b_m^2)\propto {\bm d}_m({\bm k})\times {\bm d}_m^\ast({\bm k})$ is close to the maximum value, i.e., $K_m = 1$, in the PDW phase ($ |a_m| \approx |b_m| $). 
This finding may be surprising, because the nonunitary state is unstable in purely spin-triplet superconductors.~\cite{RevModPhys.47.331,RevModPhys.63.239} 
Although the nonunitary state can be stabilized near the transition temperature through the spin-polarization effect discussed for $^{3}$He~\cite{RevModPhys.47.331} and Sr$_2$RuO$_4$~\cite{JPSJ.74.2905,JPSJ.69.1290}, the nonunitarity is suppressed at low temperatures so that the condensation energy is maximized. 
Furthermore, it has been established that spin-orbit coupling suppresses the nonunitary state~\cite{JPSJ.74.2905}. 
In our case, an inhomogeneous Rashba spin-orbit coupling gives rise to a large spin-orbit coupling of spin-triplet Cooper pairs~\cite{JPSJ.79.084701}, however, the large nonunitary component appears in the spin-triplet part as $K_m \sim 1$, owing to its special origin, which has not yet been clarified.

We show that a nonunitary spin-triplet pairing is induced by the ``field-induced parity mixing (FIPM) of Cooper pairs''. 
Because the helical-2 state belongs to a different irreducible representation of the local $C_{4v}$ point group from the $s$-wave state,~\cite{RevModPhys.63.239} parity mixing does not occur between these states at zero magnetic field, although it is allowed between the helical-1 state and the $s$-wave state. 
On the other hand, the broken time-reversal symmetry due to the magnetic field allows the mixing of the helical-1, helical-2, and $s$-wave states. 
In order to examine the magnetic field dependence of parity mixing, we calculate the irreducible superconducting susceptibility $\chi^{(0)}_{ml, m'l'}$, which is defined as 
\begin{eqnarray}
  \chi^{(0)}_{ml, m'l'}(q)=\int_0^{1/T}d\tau{\rm e}^{{\rm i}\omega_n\tau}\langle B_{ml}({\bm q},\tau)B_{m'l'}^\dagger(0)\rangle_0, 
\end{eqnarray}
with $B_{ml}^\dagger({\bm q})=(1/2)\sum_{{\bm k},\nu,s,s'}d_\nu^{\,l}({\bm k})({\rm i} \hat{\sigma}_\nu\hat{\sigma}_2)_{ss'}c^\dagger_{{\bm k}+{\bm q}ms}c^\dagger_{-{\bm k}ms'}$ being the creation operator of Cooper pairs with a center-of-mass momentum ${\bm q}$ on layer $m$.  
The average $\langle \rangle_0$ is calculated for a noninteracting Hamiltonian. 
The generalized $d$-vector including the spin-singlet component is introduced as $d^{\,l}({\bm k}) = [\psi^l({\bm k}), d_1^{\,l}({\bm k}), d_2^{\,l}({\bm k}), d_3^{\,l}({\bm k})]$.
For the $s + p$-wave superconductivity considered here, it has seven components, $d^{1}({\bm k})=(1, 0, 0, 0)$, $d^{\,2,3}({\bm k})=(0, \sin k_x, \pm \sin k_y, 0)$, $d^{\,4,5}({\bm k})=(0, \pm \sin k_y, \sin k_x, 0)$, and $d^{\,6,7}({\bm k})=(0, 0, 0, \sin k_x \pm {\rm i}\sin k_y)$. 
The mixing of the helical-2 and $s$-wave states is represented by an off-diagonal component, $\chi^{(0)}_{\rm S, H_2} \equiv \chi^{(0)}_{11,12}(0) = -\chi^{(0)}_{21,22}(0)$. 
As expected, $\chi^{(0)}_{\rm S, H_2}(0)$ vanishes at zero magnetic field, but, at high fields, its amplitude $|\chi^{(0)}_{\rm S, H_2}|$ is comparable to (or larger than) that of $\chi^{(0)}_{\rm S, H_1} \equiv \chi^{(0)}_{11,15}(0) = -\chi^{(0)}_{21,25}(0)$, representing the mixing of the helical-1 and $s$-wave states [Fig.~\ref{fig2}(b)]. 
Because $\chi^{(0)}_{\rm S, H_2}$ is purely imaginary and $\chi^{(0)}_{\rm S, H_1}$ is real, the induced spin-triplet pairing is nonunitary, as expressed in Eq.~(\ref{eq2}). 
In Fig.~\ref{fig2}(b), we also show the off-diagonal irreducible susceptibility $\chi^{(0)}_{\rm H_1, H_2} \equiv \chi^{(0)}_{15,12}(0) = \chi^{(0)}_{25,22}(0)$ causing the spin-polarization effect. 
Although its amplitude $|\chi^{(0)}_{\rm H_1, H_2}|$ is comparable to $|\chi^{(0)}_{\rm S, H_2}|$, the spin polarization effect plays a quantitatively minor role because both order parameters of the helical-1 and helical-2 states are small. 
Thus, both the helical-1 and helical-2 components of the spin-triplet order parameter are induced by parity mixing with the $s$-wave component through the Rashba spin-orbit coupling; therefore, the spin-triplet pairing is nonunitary. 
Since the condensation energy is mainly gained by spin-singlet pairing, the nonunitarity is not suppressed at low temperatures, in contrast to that in purely spin-triplet superconductors.

Although we have investigated a specific model in Eq.~(\ref{eq1}), our finding would also be valid for other models of multilayers. 
For instance, we studied the number density dependence of the superconducting phase in our model and found that the induced nonunitary spin-triplet pairing is almost independent of the number density. 
An exceptional case is that of half-filling, namely, $n=1$. 
In this case, the particle-hole symmetry prohibits the parity mixing of the helical-1 and $s$-wave states. 
Thus, the superconducting state at zero magnetic field is a purely $s$-wave BCS state, and Cooper pairs at finite magnetic fields consist of the spin-singlet $s$-wave and spin-triplet helical-2 components.

As we have shown in Fig.~\ref{fig1}, the FIPM also plays an important role in enhancing the critical magnetic field. 
When we neglect the FIPM, the critical magnetic field for $V_{\rm t}/V_{\rm s}=0.5$ depicted by the thin solid line is close to that for $V_{\rm t}/V_{\rm s} = 0$. 
This means that the enhancement of the critical magnetic field is mainly caused through the FIPM.

\begin{figure}[htbp]
  \begin{center}
    \begin{tabular}{cc}
      \begin{minipage}{0.5\hsize}
        \includegraphics[width=40mm]{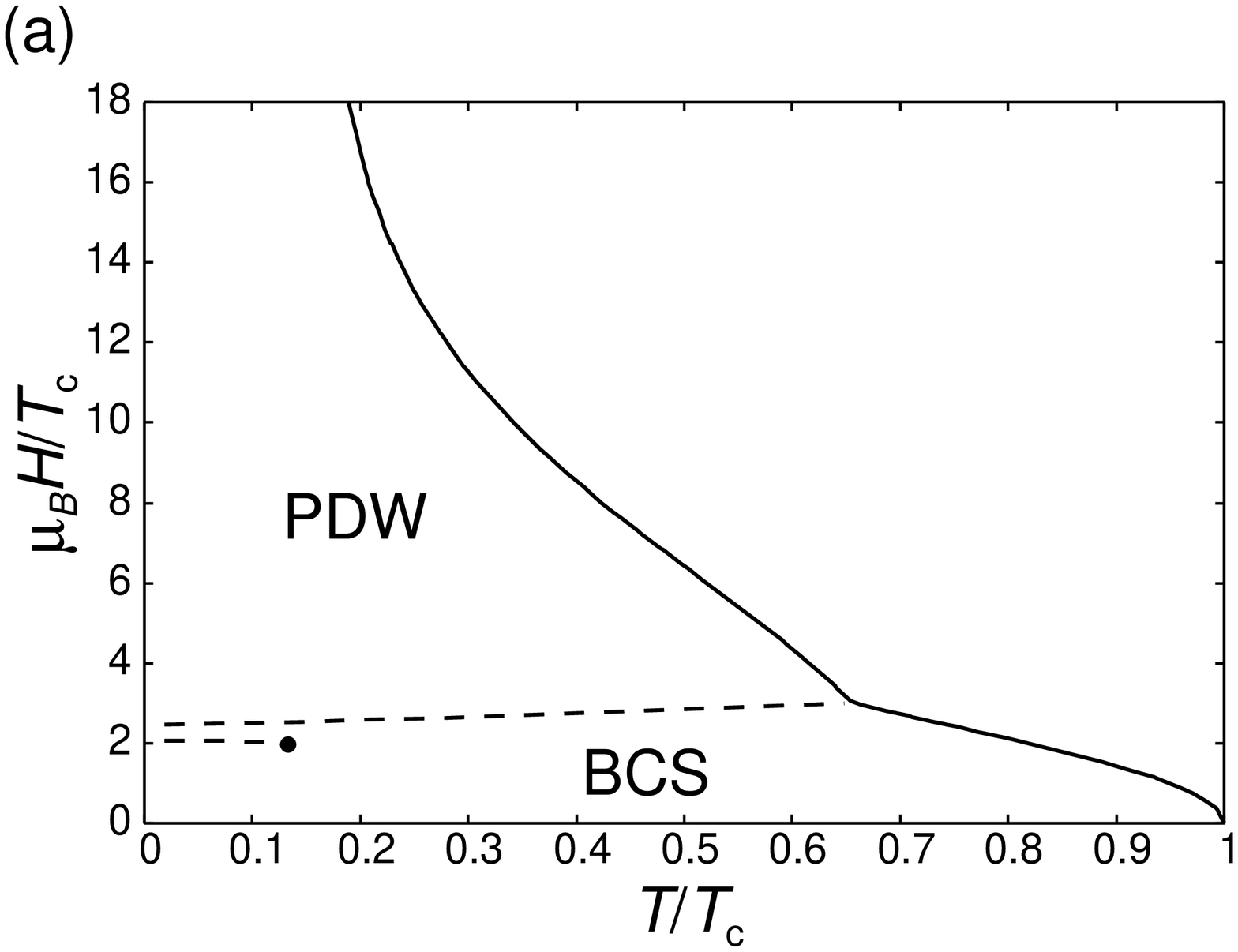}
      \end{minipage}
      \begin{minipage}{0.5\hsize}
        \includegraphics[width=40mm]{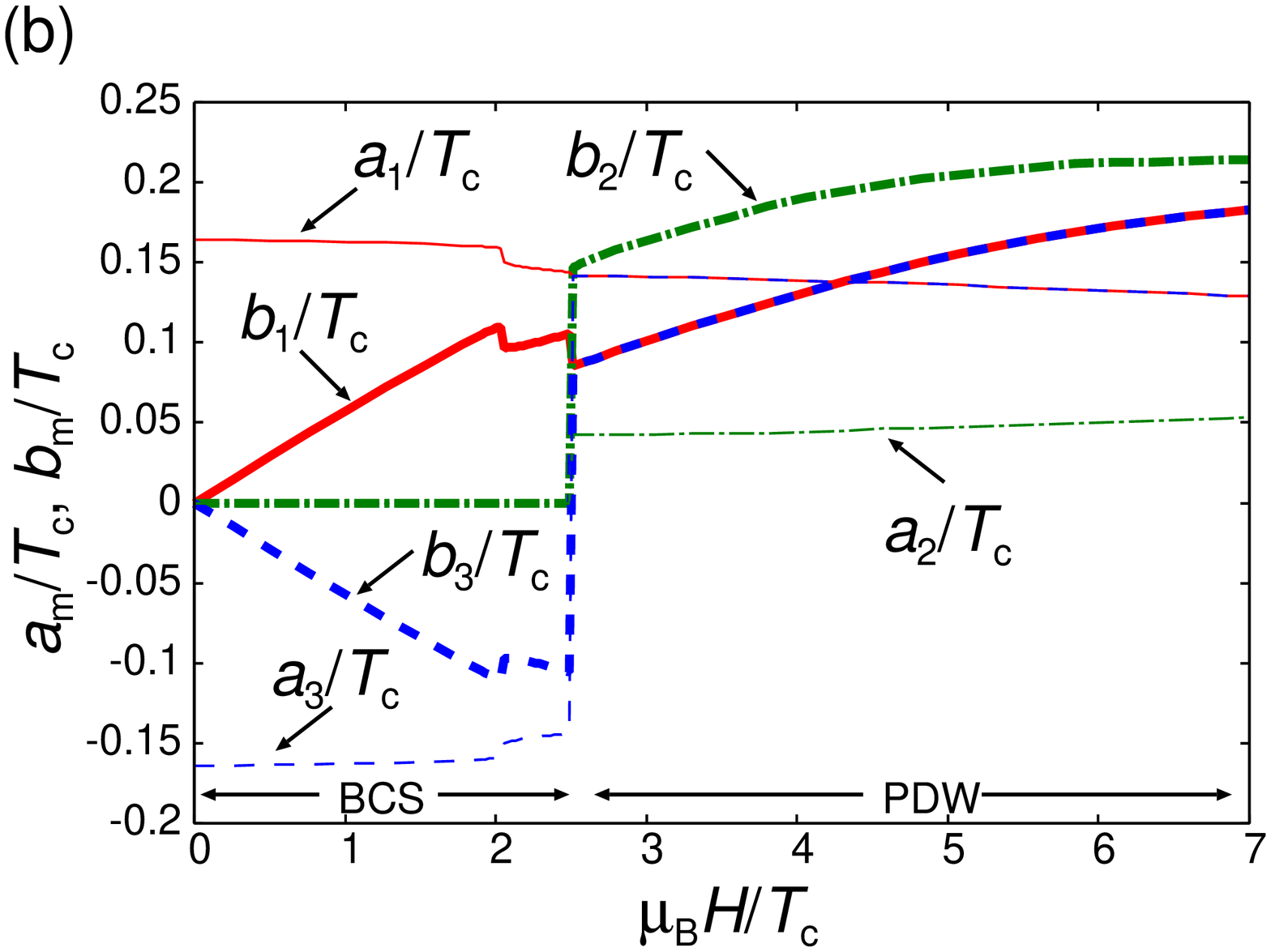}
      \end{minipage}
    \end{tabular}
    \caption{(Color online) (a) $T$-$H$ phase diagram in the trilayer system for $V_{\rm t}/V_{\rm s} = 0.5$. 
      The solid and dashed lines show the second-order and first-order phase transition lines, respectively. 
      The black dot shows the tricritical point.
      (b) Magnetic field dependence of the $d$-vector in the trilayer system at the low temperature $T/T_{\rm c}=0.0953$.
      The temperature and order parameters are normalized by the transition temperature at zero magnetic field with $T_{\rm c}=0.0262$. 
    }
    \label{fig3}
  \end{center}
\end{figure}
Now we turn to the trilayer systems. We see a unique property that is absent in the bilayer system: the existence of an inner layer that even has the local inversion symmetry. 
The order parameters show the layer dependence in the BCS state as $(\psi_1, \psi_2, \psi_3)=(\psi_{\rm out}, \psi_{\rm in}, \psi_{\rm out})$ and $({\bm d}_1, {\bm d}_2, {\bm d}_3)=({\bm d}_{\rm out}, {\bm 0}, -{\bm d}_{\rm out})$, while those in the PDW state are $(\psi_1, \psi_2, \psi_3)=(\psi_{\rm out}, 0, -\psi_{\rm out})$ and $({\bm d}_1, {\bm d}_2, {\bm d}_3)=({\bm d}_{\rm out}, {\bm d}_{\rm in}, {\bm d}_{\rm out})$. 
The $T$-$H$ phase diagram is similar to that of bilayer systems, except for the first-order phase transition line with a tricritical point in the BCS state [Fig.~\ref{fig3}(a)]. 
In this first-order phase transition, the order parameter in the inner layer is discontinuously suppressed by the paramagnetic depairing effect, as we elucidated for $V_{\rm t}/V_{\rm s}=0$~\cite{PRB.86.134514}.
Thus, the phase diagram is not affected by the parity mixing of Cooper pairs except for the enhancement of the critical magnetic field.

Moreover, we see an intriguing property in the induced spin-triplet pairing. 
The $d$-vector is again described by Eq.~(\ref{eq2}), and Fig.~\ref{fig3}(b) shows the magnetic field dependence of the coefficients $a_m$ and $b_m$. 
The $d$-vector in the outer layers shows a similar field dependence to that in bilayers, i.e., nonunitary spin-triplet pairing. 
On the other hand, the induced spin-triplet pairing in the inner layer is almost unitary, as indicated by $|a_2| \ll |b_2|$.  
Because the Rashba spin-orbit coupling vanishes at the inner layer that conserved even the local inversion symmetry, the helical-2 state is negligibly suppressed by the spin-orbit coupling, and therefore is the main component in the spin-triplet channel.

\begin{figure}[htbp]
  \begin{center}
    \begin{tabular}{cc}
      \begin{minipage}{0.5\hsize}
        \includegraphics[width=40mm]{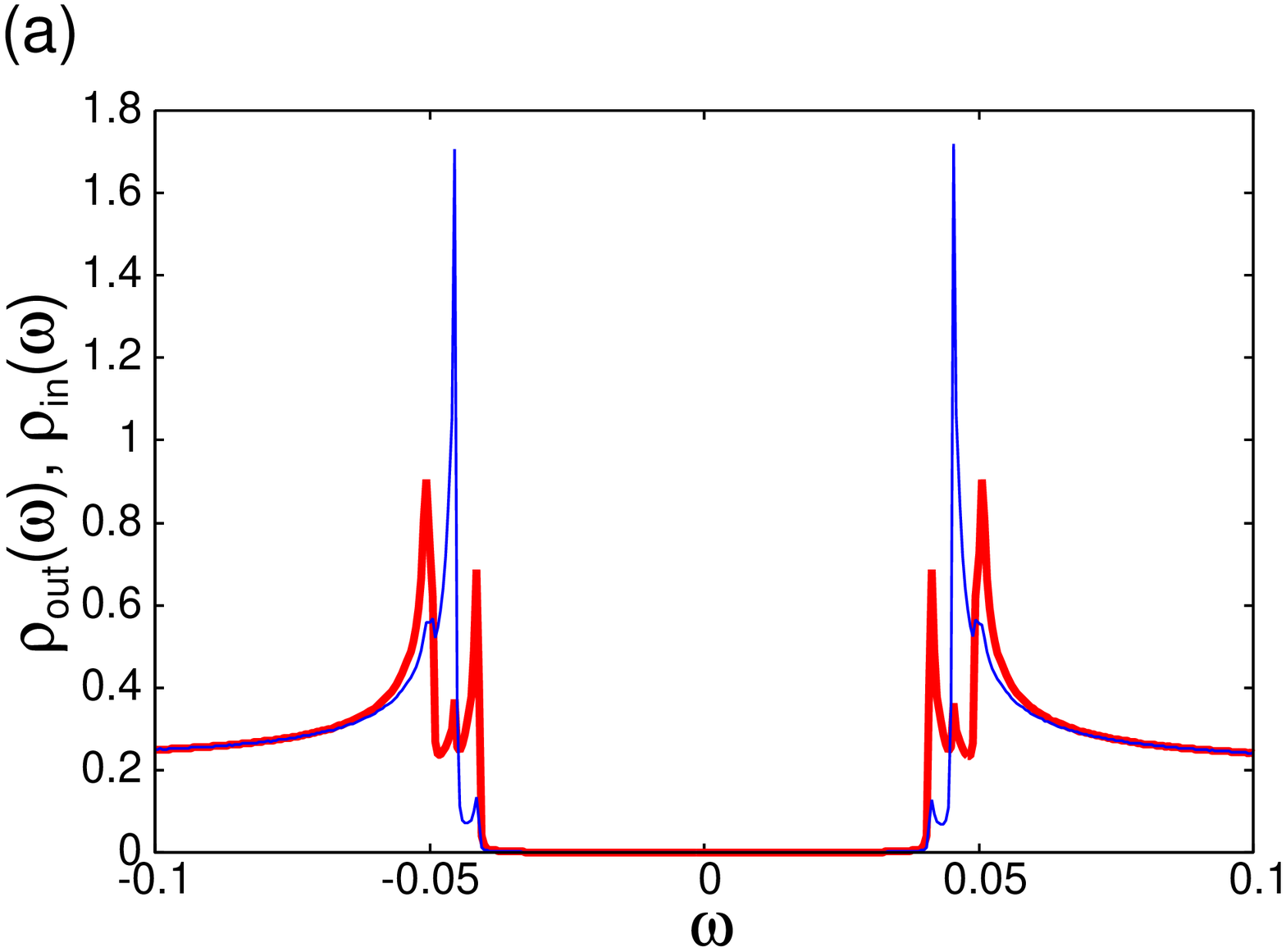}
      \end{minipage}
      \begin{minipage}{0.5\hsize}
        \includegraphics[width=40mm]{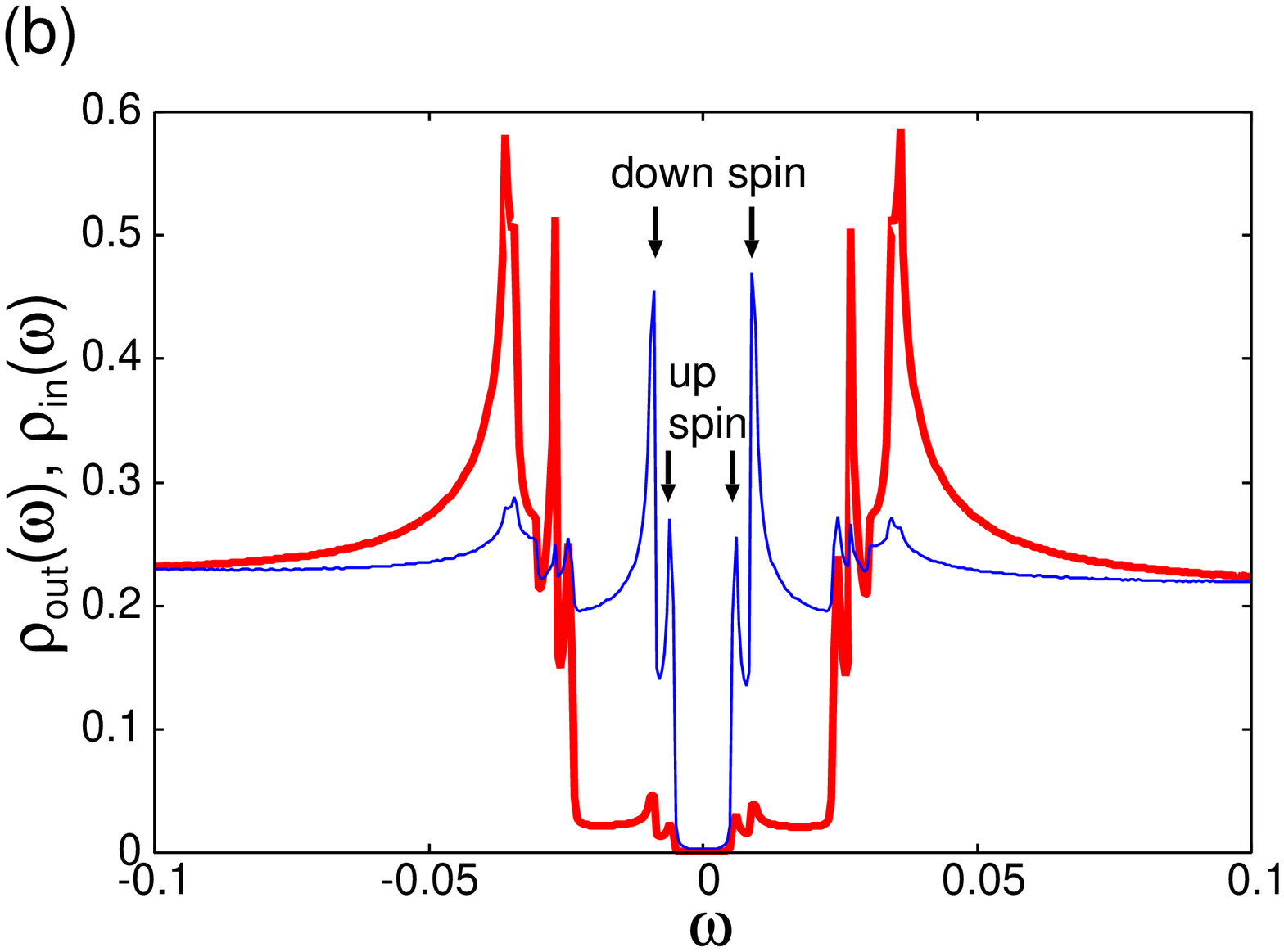}
      \end{minipage}
    \end{tabular}
    \caption{(Color online) LDOS in the trilayer system at $T/T_{\rm c}=0.0953$. 
      The thin (blue) and thick (red) lines show the LDOS at the inner layer $\rho_{\rm in}(\omega)$ and that at the outer layer $\rho_{\rm out}(\omega)$, respectively.  
      (a) $\mu_{\rm B}H/T_{\rm c}=0$ (BCS phase) and (b) $\mu_{\rm B}H/T_{\rm c}=3$ (PDW phase). 
      Quasiparticles with up spin and those with down spin give rise to the peaks of LDOS, as shown in Fig.~\ref{fig4}(b). 
      The other parameters are the same as those in Fig.~\ref{fig3}.
    }
    \label{fig4}
  \end{center}
\end{figure}
We here propose an experimental test on the parity-mixed superconducting phase in the LNCS. 
We focus on the trilayer system, since a recent experiment obtained evidence of a staggered Rashba spin-orbit coupling in the trilayer superlattice of CeCoIn$_5$~\cite{PRL.109.157006}. 
Figure~\ref{fig4} shows the local density of states (LDOS) obtained by solving the BdG equation. 
In the BCS phase, the LDOS shows a three-gap structure; one peak of LDOS mainly comes from the inner layer and the other two peaks come from the outer layers [Fig.~\ref{fig4}(a)]. 
To understand this three-gap structure, we show the order parameter in the band basis, which has been obtained from Ref.~\citen{JPSJ.81.034702}: 
\begin{eqnarray}
  \Delta_1&=&\psi+\frac{\alpha|{\bm g}({\bm k})|^2}{\sqrt{\alpha^2|{\bm g}({\bm k})|^2+2t_\perp^2}}d, \\
  \Delta_2&=&\psi, \\
  \Delta_3&=&\psi-\frac{\alpha|{\bm g}({\bm k})|^2}{\sqrt{\alpha^2|{\bm g}({\bm k})|^2+2t_\perp^2}}d,
\end{eqnarray}
where we assume $\psi_{\rm out}=\psi_{\rm in}=\psi$ and ${\bm d}_{\rm out}({\bm k})=d{\bm g}({\bm k})$. 
If we neglect parity mixing, namely, $d=0$, these order parameters are equivalent; therefore, the LDOS shows a single-gap structure, as we have shown in Ref.~\citen{PRB.86.134514}.  
Thus, the three-gap structure in the LDOS is a signature of the parity-mixed superconducting state. 
This gap structure is distinguished from the two-gap structure arising from the imbalance of the inner and outer layers, namely, $\psi_{\rm out} \ne \psi_{\rm in}$.

A characteristic property of the PDW state appears in the pronounced multigap structure, as shown in Fig.~\ref{fig4}(b). 
As expected, the gap is reduced in the inner layer. 
A signature of the nonunitary spin-triplet pairing in the PDW phase is observed in the subgap structure of the inner layer. 
Although we obtain a single small gap without taking parity mixing into account~\cite{PRB.86.134514}, the LDOS in the inner layer shows two small gaps, as shown in Fig.~\ref{fig4}(b). 
One is the gap of quasiparticles having up spins, and the other is the gap of down spins. 
This subgap structure arises from the nonunitary spin-triplet pairing and is not due to the Zeeman shift. 
Indeed, the Zeeman shift of Bogoliubov quasiparticles does not occur in the PDW phase~\cite{JPSJ.81.034702}.

Finally, we comment on several points. 
First, we focused on the dominant spin-singlet superconductivity, because most of the superconductors belong to this class. 
When we assume a large pairing interaction in the spin-triplet channel, $V_{\rm t}$, the dominant spin-triplet pairing state is stabilized independent of the magnetic field. 
This is a topologically nontrivial state, as shown in Refs.~\citen{PRB.79.060505} and \citen{PRB.79.094504}. 
The dominant spin-singlet pairing state studied here also has a topologically nontrivial property, as shown in Ref.~\citen{PRL.108.147003} for bilayers. 
Basically, the odd parity PDW state can be a topological superconducting state. 
We will show the topological properties of the PDW state in more than three layers in another report.

Next, we comment on another exotic superconducting phase induced by the frustration of spin-singlet pairing and spin-triplet pairing.
In the multilayer superconductors, the Josephson coupling energy is gained through the spin-singlet pairing in the BCS phase, while the spin-triplet pairing favors the PDW phase. 
When $T_{\rm c}^{\rm s}\sim T_{\rm c}^{\rm t}$, frustration occurs and gives rise to a fractional phase difference in both the spin-singlet and spin-triplet order parameters between layers. 
Then, the superconducting state is accompanied by a spontaneous time-reversal symmetry breaking. 
This situation is similar to that in non-centrosymmetric superconductors having a twin boundary.~\cite{JPSJ.77.083701,PRB.87.220504} 
We find that such a superconducting phase is stabilized when we assume $V_{\rm t}/V_{\rm s} \sim 1$. 
This superconducting phase may be realized in the locally non-centrosymmetric superconductor SrPtAs, in which a recent experiment showed a spontaneous time-reversal symmetry breaking in the superconducting state.~\cite{PRB.87.180503}

In summary, we have investigated the parity-mixed superconductivity arising from the local violation of inversion symmetry in multilayer systems. 
At zero magnetic field, a staggered spin-triplet superconductivity is induced by a uniform spin-singlet superconductivity. 
The signature of this superconducting state appears in the characteristic multigap structure which can, in principle, be tested by experiments. 
More interestingly, the uniform spin-triplet superconductivity is induced in the PDW phase, which is stabilized in magnetic fields along the [001]-axis. 
In other words, the odd-parity superconductivity is stabilized by the synergistic roles of local inversion symmetry breaking and magnetic fields. 
We have shown that the spin-triplet superconductivity induced by this manner is nonunitary owing to the FIPM. 
Importantly, we do not need a strong attractive interaction in the spin-triplet channel for this mechanism of spin-triplet superconductivity.
This result helps to design spin-triplet superconductors, which are attracting increasing attention~\cite{JPSJ.81.011013} but rarely realized in real materials. 
Since the PDW state is suppressed by the orbital depairing effect, we should study Pauli-limited superconductors with a large Maki parameter. 
Concerning this point, the artificial superlattice of CeCoIn$_5$~\cite{NatPhys.7.849,PRL.109.157006} is one of the preferable systems. 
It is highly desirable to clarify this superconducting phase and design another exotic superconducting phase in the LNCS.

\section*{Acknowledgements} 
We are grateful to S.~K.~Goh, D.~Maruyama, Y.~Matsuda, T.~Shibauchi, M.~Shimozawa, and H.~Shishido for fruitful discussions. 
This work was supported by KAKENHI (Grant Nos.~25103711, 24740230, and 23102709). 
We are also grateful for the financial support from the Swiss Nationalfonds, the NCCR MaNEP, and the Pauli Center of ETH Zurich.


\begin{thebibliography}{10}

\bibitem{Science.301.1348}
S.~Murakami, N.~Nagaosa, and S.-C.~Zhang: Science {\bfseries 301} (2003) 1348.

\bibitem{PRL.92.126603}
J.~Sinova, D.~Culcer, Q.~Niu, N.~A.~Sinitsyn, T.~Jungwirth, and A.~H.~MacDonald: Phys. Rev. Lett. {\bfseries 92} (2004) 126603.

\bibitem{Nature.465.901}
X.~Z.~Yu, Y.~Onose, N.~Kanazawa, J.~H.~Park, J.~H.~Han, Y.~Matsui, N.~Nagaosa, and Y.~Tokura: Nature {\bfseries 465} (2010) 901.

\bibitem{PRL.108.107202}
Y.~Togawa, T.~Koyama, K.~Takayanagi, S.~Mori, Y.~Kousaka, J.~Akimitsu, S.~Nishihara, K.~Inoue, A.~S.~Ovchinnikov, and J.~Kishine: 
Phys. Rev. Lett. {\bfseries 108} (2012) 107202.

\bibitem{RevModPhys.82.3045}
M.~Z. Hasan and C.~L. Kane: Rev. Mod. Phys. {\bfseries 82} (2010) 3045.

\bibitem{RevModPhys.83.1057}
X.-L.~Qi and S.-C.~Zhang: Rev. Mod. Phys. {\bfseries 83} (2011) 1057.

\bibitem{NCSC}
E.~Bauer and M.~Sigrist, {\em Non-Centrosymmetric Superconductors: Introduction and Overview (Lecture Notes in Physics)}, (Springer, Berlin/Heidelberg, 2012).

\bibitem{JPSJ.79.084701}
Y.~Yanase: J. Phys. Soc. Jpn. {\bfseries 79} (2010) 084701.

\bibitem{PRB.84.184533}
M.~H. Fischer, F.~Loder, and M.~Sigrist: Phys. Rev. B {\bfseries 84} (2011) 184533.

\bibitem{JPSJ.81.034702}
D.~Maruyama, M.~Sigrist, and Y.~Yanase: J. Phys. Soc. Jpn. {\bfseries 81} (2012) 034702.

\bibitem{PRB.85.220505}
S.~J.~Youn, M.~H.~Fischer, S.~H.~Rhim, M.~Sigrist, and D.~F.~Agterberg: Phys. Rev. B {\bfseries 85} (2012) 220505.

\bibitem{PRB.86.100507}
J.~Goryo, M.~H.~Fischer, and M.~Sigrist: Phys. Rev. B {\bfseries 86} (2012) 100507.

\bibitem{PRB.86.134514}
T.~Yoshida, M.~Sigrist, and Y.~Yanase: Phys. Rev. B {\bfseries 86} (2012) 134514.

\bibitem{JPSJ.82.074714}
T.~Yoshida, M.~Sigrist, and Y.~Yanase: J. Phys. Soc. Jpn. {\bfseries 82} (2013) 074714.

\bibitem{NatPhys.7.849}
Y.~Mizukami, H.~Shishido, T.~Shibauchi, M.~Shimozawa, S.~Yasumoto, D.~Watanabe, M.~Yamashita, H.~Ikeda, T.~Terashima, H.~Kontani, and Y.~Matsuda: 
Nat. Phys. {\bfseries 7} (2011) 849.

\bibitem{PRL.96.087001}
H.~Mukuda, M.~Abe, Y.~Araki, Y.~Kitaoka, K.~Tokiwa, T.~Watanabe, A.~Iyo, H.~Kito, and Y.~Tanaka: Phys. Rev. Lett. {\bfseries 96} (2006) 087001; 
H.~Mukuda, S.~Shimizu, A.~Iyo, and Y.~Kitaoka: J. Phys. Soc. Jpn. {\bfseries 81} (2012) 011008.

\bibitem{RevModPhys.47.331}
A.~J.~Leggett: Rev. Mod. Phys. {\bfseries 47} (1975) 331.

\bibitem{RevModPhys.63.239}
M.~Sigrist and K.~Ueda: Rev. Mod. Phys. {\bfseries 63} (1991) 239.

\bibitem{SovPhysSolidState.1.368}
E.~I.~Rashba: Sov. Phys. Solid State {\bfseries 1} (1959) 368.

\bibitem{PRL.102.207004}
D.~F.~Agterberg, M.~Sigrist, and H.~Tsunetsugu: Phys. Rev. Lett. {\bfseries 102} (2009) 207004.

\bibitem{PRL.92.097001}
P.~A.~Frigeri, D.~F.~Agterberg, A.~Koga, and M.~Sigrist: Phys. Rev. Lett. {\bfseries 92} (2004) 097001.

\bibitem{JPSJ.76.051008}
S.~Fujimoto: J. Phys. Soc. Jpn. {\bfseries 76} (2007) 051008.

\bibitem{JPSJ.76.124709}
Y.~Yanase and M.~Sigrist: J. Phys. Soc. Jpn. {\bfseries 76} (2007) 124709.

\bibitem{JPSJ.74.2905}
M.~Udagawa, Y.~Yanase, and M.~Ogata: J. Phys. Soc. Jpn. {\bfseries 74} (2005) 2905.

\bibitem{JPSJ.69.1290}
M.~Sigrist: J. Phys. Soc. Jpn. {\bfseries 69} (2000) 1290.

\bibitem{PRL.109.157006}
S.~K.~Goh, Y.~Mizukami, H.~Shishido, D.~Watanabe, S.~Yasumoto, M.~Shimozawa, M.~Yamashita, T.~Terashima, Y.~Yanase, T.~Shibauchi, A.~I. Buzdin, and Y.~Matsuda: 
Phys. Rev. Lett. {\bfseries 109} (2012) 157006.

\bibitem{PRB.79.060505}
Y.~Tanaka, T.~Yokoyama, A.~V.~Balatsky, and N.~Nagaosa: Phys. Rev. B {\bfseries 79} (2009) 060505.

\bibitem{PRB.79.094504}
M.~Sato and S.~Fujimoto: Phys. Rev. B {\bfseries 79} (2009) 094504.

\bibitem{PRL.108.147003}
S.~Nakosai, Y.~Tanaka, and N.~Nagaosa: Phys. Rev. Lett. {\bfseries 108} (2012) 147003.

\bibitem{JPSJ.77.083701}
C.~Iniotakis, S.~Fujimoto, and M.~Sigrist: J. Phys. Soc. Jpn. {\bfseries 77} (2008) 083701.

\bibitem{PRB.87.220504}
E.~Arahata, T.~Neupert, and M.~Sigrist: Phys. Rev. B {\bfseries 87} (2013) 220504.

\bibitem{PRB.87.180503}
P.~K. Biswas, H.~Luetkens, T.~Neupert, T.~St\"urzer, C.~Baines, G.~Pascua, A.~P.~Schnyder, M.~H. Fischer, J.~Goryo, M.~R.~Lees, H.~Maeter, F.~Br\"uckner, H.-H.~Klauss, M.~Nicklas, P.~J.~Baker, A.~D.~Hillier, M.~Sigrist, A.~Amato, and D.~Johrendt: Phys. Rev. B {\bfseries 87} (2013) 180503.

\bibitem{JPSJ.81.011013}
Y.~Tanaka, M.~Sato, and N.~Nagaosa: J. Phys. Soc. Jpn. {\bfseries 81} (2012) 011013.
\end{thebibliography}
\end{document}